\documentclass[twoside,twocolumn,9pt]{article}
\usepackage{lipsum}   
\usepackage{extsizes}
\usepackage[super,sort&compress,comma]{natbib} 
\usepackage[version=3]{mhchem}
\usepackage[left=1.5cm, right=1.5cm, top=1.785cm, bottom=2.0cm]{geometry}
\usepackage{balance}
\usepackage{mathptmx}
\usepackage{sectsty}
\usepackage{graphicx} 
\graphicspath{{./Figures/}{./}}
\usepackage{lastpage}
\usepackage[format=plain,justification=justified,singlelinecheck=false,font={stretch=1.125,small,sf},labelfont=bf,labelsep=space]{caption}
\usepackage{float}
\usepackage{fancyhdr}
\usepackage{fnpos}
\usepackage[english]{babel}
\addto{\captionsenglish}{%
  
}
\usepackage{array}
\usepackage{droidsans}
\usepackage{charter}
\usepackage[T1]{fontenc}
\usepackage[usenames,dvipsnames]{xcolor}
\usepackage{setspace}
\usepackage[compact]{titlesec}
\usepackage{hyperref}

\definecolor{cream}{RGB}{222,217,201}

\begin{document}

\pagestyle{fancy}
\thispagestyle{plain}
\fancypagestyle{plain}{
\renewcommand{\headrulewidth}{0pt}
}

\makeFNbottom
\makeatletter
\renewcommand\LARGE{\@setfontsize\LARGE{15pt}{17}}
\renewcommand\Large{\@setfontsize\Large{12pt}{14}}
\renewcommand\large{\@setfontsize\large{10pt}{12}}
\renewcommand\footnotesize{\@setfontsize\footnotesize{7pt}{10}}
\makeatother

\renewcommand{\thefootnote}{\fnsymbol{footnote}}
\renewcommand\footnoterule{\vspace*{1pt}%
\color{cream}\hrule width 3.5in height 0.4pt \color{black}\vspace*{5pt}} 
\setcounter{secnumdepth}{5}

\makeatletter 
\renewcommand\@biblabel[1]{#1}            
\renewcommand\@makefntext[1]%
{\noindent\makebox[0pt][r]{\@thefnmark\,}#1}
\makeatother 
\renewcommand{\figurename}{\small{Fig.}~}
\sectionfont{\sffamily\Large}
\subsectionfont{\normalsize}
\subsubsectionfont{\bf}
\setstretch{1.125} 
\setlength{\skip\footins}{0.8cm}
\setlength{\footnotesep}{0.25cm}
\setlength{\jot}{10pt}
\titlespacing*{\section}{0pt}{4pt}{4pt}
\titlespacing*{\subsection}{0pt}{15pt}{1pt}

\fancyfoot{}
\fancyfoot[LO,RE]{\vspace{-7.1pt}\includegraphics[height=9pt]{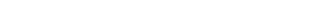}}
\fancyfoot[CO]{\vspace{-7.1pt}\hspace{13.2cm}\includegraphics{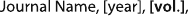}}
\fancyfoot[CE]{\vspace{-7.2pt}\hspace{-14.2cm}\includegraphics{head_foot/RF}}
\fancyfoot[RO]{\footnotesize{\sffamily{1--\pageref{LastPage} ~\textbar  \hspace{2pt}\thepage}}}
\fancyfoot[LE]{\footnotesize{\sffamily{\thepage~\textbar\hspace{3.45cm} 1--\pageref{LastPage}}}}
\fancyhead{}
\renewcommand{\headrulewidth}{0pt} 
\renewcommand{\footrulewidth}{0pt}
\setlength{\arrayrulewidth}{1pt}
\setlength{\columnsep}{6.5mm}
\setlength\bibsep{1pt}

\makeatletter 
\newlength{\figrulesep} 
\setlength{\figrulesep}{0.5\textfloatsep} 

\newcommand{\topfigrule}{\vspace*{-1pt}%
\noindent{\color{cream}\rule[-\figrulesep]{\columnwidth}{1.5pt}} }

\newcommand{\botfigrule}{\vspace*{-2pt}%
\noindent{\color{cream}\rule[\figrulesep]{\columnwidth}{1.5pt}} }

\newcommand{\dblfigrule}{\vspace*{-1pt}%
\noindent{\color{cream}\rule[-\figrulesep]{\textwidth}{1.5pt}} }

\makeatother

\twocolumn[
  \begin{@twocolumnfalse}
{\includegraphics[height=30pt]{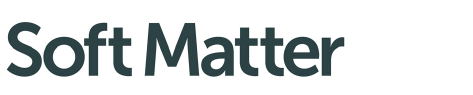}\hfill\raisebox{0pt}[0pt][0pt]{\includegraphics[height=55pt]{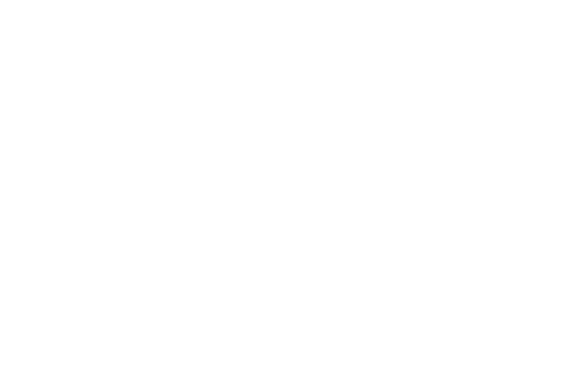}}\\[1ex]
\includegraphics[width=18.5cm]{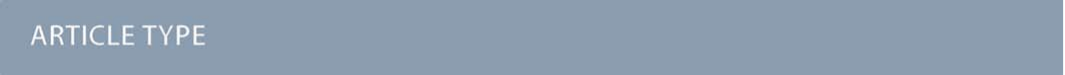}}\par
\vspace{1em}
\sffamily
\begin{tabular}{m{4.5cm} p{13.5cm} }

\includegraphics{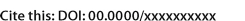} & \noindent\LARGE{\textbf{Banded phases in topological flocks}} \\
\vspace{0.3cm} & \vspace{0.3cm} \\

 & \noindent\large{Charles R. Packard,\textit{$^{a,\ddag}$} and Daniel M. Sussman\textit{$^{a,\ast}$}} \\

\includegraphics{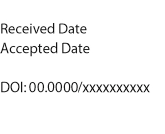} & \noindent\normalsize{
    Flocking phase transitions found in models of polar active matter are paradigmatic examples of active phase transitions in soft matter.
    An interesting specialization of flocking models concerns a ``topological'' vs ``metric'' choice by which agents are considered to be interacting neighbors.
    While recent theoretical work suggests that the order-disorder transition in these polar aligning models is universally first order, numerical studies have suggested that topological models may instead have a continuous transition.
    Some recent simulations have found that some variations of topologically interacting flocking agents have a discontinuous transition, but unambiguous observations of phase coexistence using common Voronoi-based alignment remains elusive.
    In this work, we use a custom GPU-accelerated simulation package to perform million-particle-scale simulations of these Voronoi-Vicsek flocking models.
    By accessing such large systems on appropriately long time scales, we are able to show that a regime of stable phase coexistence between the ordered and disordered phases, confirming the discontinuous nature of this transition in the thermodynamic limit.
} \\

\end{tabular}

 \end{@twocolumnfalse} \vspace{0.6cm}

  ]

\renewcommand*\rmdefault{bch}\normalfont\upshape
\rmfamily
\section*{}
\vspace{-1cm}


\footnotetext{\textit{$^{a}$~Department of Physics, Emory University, Atlanta, Georgia 30322, USA }}
\footnotetext{$^{\ddag}$~Email: charles.robert.packard@emory.edu}
\footnotetext{$^{\ast}$~Email: daniel.m.sussman@emory.edu}



\section{Introduction}

The spontaneous self-organization of synthetic or biological self-propelled agents into a state of ordered collective motion is observed in nature from microscopic \cite{chen2017weak} to macroscopic \cite{bazazi2008collective,sugi2019c,cavagna2010scale} length-scales.
Despite the complexity of their constituent components, many of the emergent large-scale dynamics in these experimental systems can be understood from highly coarse-grained agent-based models \cite{sumino2012large,chen2017weak,xu2024self}.
The Vicsek model \cite{vicsek1995novel} -- a foundational model in the study of active matter -- describes ``flocks'' of polar aligning agents.
A substantial body of research has focused on understanding the nature of the phase transition between the ordered and disordered states of this model (and its many variants) \cite{nagai2015collective,zhao2021phases}, as well as the hydrodynamic properties of systems with the same symmetries \cite{toner2005hydrodynamics}.

In its original formulation, agents in the Vicsek model align with all agents within a characteristic length scale \cite{chate2008collective}.
This ``metric'' flavor of the model is particularly well-suited for flocks such as active colloids \cite{bricard2013emergence}, microtubules \cite{sumino2012large}, and bacteria \cite{chen2017weak}, in which aligning interactions between agents occurs via collisions.
In macroscopic flocks though (e.g., some insect \cite{buhl2011group}, fish \cite{lopez2012behavioural}, and bird flocks~\cite{ballerini2008interaction}), alignment interactions may be vision-based or otherwise ``metric-free'' \cite{ginelli2010relevance} or ``topological''.
That is, the network of neighbor interactions stems not directly from a pairwise distance, but from some other criteria (for instance, $k$-nearest-neighbors or from a Voronoi tessellation of space).
The question of whether or not metric and topological flocks have order-disorder transitions in the same universality class remains an active source of discussion \cite{martin2021fluctuation,martin2024fluctuation}.

In metric flocks, the transition to collective motion is discontinuous \cite{chate2008collective}, with phase coexistence at the phase boundary that is understood as a non-equilibrium analogue of the liquid-gas transition \cite{solon2015phase}.
This phase behavior has been studied using Boltzmann-style coarse-graining methods which allow the dynamics of flocks to be treated at the level of interacting fields \cite{bertin2006boltzmann}.
The resulting hydrodynamics reveal that the microscopic length-scale of aligning interactions in metric flocks leads to a coupling between the coarse-grained density and polarization fields, which subsequently produces a linear instability of the homogeneous ordered phase in the vicinity of the order-disorder phase boundary and gives rise to phase coexistence \cite{solon2015pattern}.
This phase coexistence is characterized by the presence of high-density, highly-ordered ``bands'' of particles propagating through a low-density, disordered background of particles.
This phase transition is well-known in metric Vicsek models, and has been experimentally observed in both synthetic \cite{bricard2013emergence} and biological flocks \cite{buhl2011group}.

A field-theoretic analysis of topological flocks suggested that the absence of a microscopic length-scale correspondingly results in the absence of a coupling between the density and polarization fields \cite{peshkov2012continuous,chou2012kinetic} -- this would rationalize the continuous order-disorder transition previously reported for the Voronoi-Vicsek model \cite{ginelli2010relevance}.
Recently, though, theoretical work has emerged suggesting that correlated fluctuations in the coarse-grained density and velocity fields of a topological flock lead to renormalized hydrodynamics that make the phase transition \emph{discontinuous} \cite{martin2021fluctuation}.
At present, this fluctuation-induced first-order transition scenario~\cite{binder1987theory} has been demonstrated in agent-based simulations of the topological $k$-nearest neighbor active Ising model (AIM) \cite{martin2021fluctuation} -- a lattice flocking model with discrete symmetry.
Preliminary evidence suggests the same discontinuous transition characterizes the Voronoi variant of the AIM \cite{martin2024fluctuation}, albeit at the very edge of system size that the computational methods employed in that work could resolve.
These studies on topological AIM flocks are further complicated by the fact that their active orientational dynamics have a discrete $\mathcal{Z}_2$ symmetry, in contrast to the continuous rotational symmetry of both Vicsek and natural flocks.
Recent work has found that such differences can have profound effects on the macroscopic phase behavior of the ordered states \cite{karmakar2024consequence}.
In addition to the unclear theoretical scenario, this highlights a tension in the numerical literature between the recent AIM models and previous simulations of Vicsek-style Voronoi flocks which, as mentioned before, found no signature of phase coexistence in a thorough finite-size scaling analysis \cite{ginelli2010relevance}.

In this paper, we conduct large-scale numerical simulations to resolve this tension.
We first show that the discrepancy between earlier \cite{ginelli2010relevance} and more recent \cite{martin2024fluctuation} work vanishes when one considers sufficiently large systems and takes the time-continuous limit of a Voronoi-Vicsek model.
Standard Vicsek model simulations use discrete-time dynamics (in which positions and orientations are updated with a time-step size of $\Delta t = 1$), in which arbitrarily large changes in particle orientation are permitted~\cite{vicsek1995novel}.
We adopt a time-continuous formulation in which the equations of motion are expressed as coupled differential equations (and for which angular updates are controlled by a conservative potential), allowing us to independently vary particle speeds and the discretization of time in our simulations.
We perform a finite-size analysis to demonstrate that the order-disorder transition in Voronoi flocks is discontinuous.
We furthermore expand on the results of Ref.~\cite{martin2024fluctuation} by explicitly demonstrating coexisting phases with a bimodal distribution of densities, and we directly measure a non-vanishing coupling between density and polarization fields.

In the remainder of this manuscript, we first describe the model and our numerical methods in more detail.
We then report coarse-grained field statistics for our large-scale simulations, discussing the structure of the coexisting phases and comparing the traditional discrete-time version of the model with a time-continuous implementation.
We finally close with a discussion and outlook on questions of flocking in metric and topologically-interacting systems.

\section{Models and methods}

\begin{figure}[h]
	\centering
    \includegraphics[width=0.99\columnwidth]{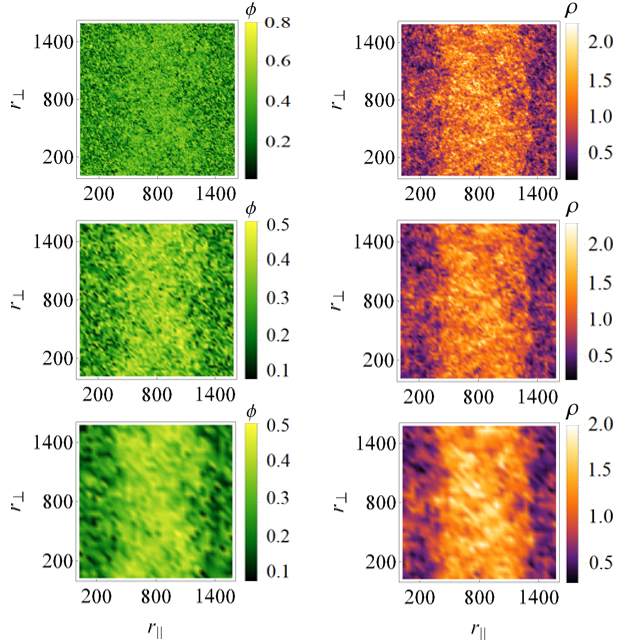}
    \caption{\textbf{Coarse-grained field snapshots.}
		Instantaneous order parameter (left) and density (right) field configurations for simulations of $N=2.56\times10^6$ particles. The noise strength is set to $\eta=0.4880$. Fields are constructed by coarse-graining with boxes of size $\ell=12.5$ (top), $\ell=25.0$ (middle),  and $\ell=50.0$ (bottom).
		The coordinate system is chosen such that the x-axis is parallel to the mean flocking direction.
	}
	\label{fig:CoarseGrainedFieldSnapshots}
\end{figure}

The original Vicsek model considered discrete-time updates of particle positions and orientations, and furthermore used an angular update whose action cannot be expressed as the derivative of a potential \cite{chepizhko2021revisiting}.
We adopt a numerical model more amenable to studying time-continuous dynamics, and also focus on a ``reciprocal'' version of the Vicsek model, in which the particle torques are governed by a classical XY-model Hamiltonian.
Our simulations are of $N$ particles in a 2D square box of linear size $L$ with periodic boundary conditions.
The linear size of the simulation domain reflects of our unit of length, and without loss of generality we fix the particle density at $\rho_{0}=N/L^2=1.0$.
The positions and orientations of particles evolve in time according to
\begin{equation}
\label{eq::microscopic_pos_update}
\frac{d\mathbf{r}_{i}(t)}{dt} = v_{0}\begin{bmatrix}
\cos\theta_{i}(t) \\ \sin\theta_{i}(t)
\end{bmatrix},
\end{equation}
\begin{equation}
\label{eq::microscopic_theta_update}
\frac{d\theta_{i}(t)}{dt} = -\nabla_{\theta_{i}}H + \eta\,\zeta_{i}(t)\,.
\end{equation}
Here $v_{0}$ is the self-propulsion speed, and the energy of particle $i$ is given by
\begin{equation}
\label{eq::microscopic_Hamiltonian}
H[\{\theta_{i}(t)\}]=\alpha\sum_{{j}\in\mathcal{N}_i(t)}\cos\left[\theta_{j}(t)-\theta_{i}(t)\right]\,.
\end{equation}
The parameter $\alpha$ sets the interaction strength of the polar alignment.

In the standard Vicsek model $\mathcal{N}_i(t)$ (the set of neighbors of particle $i$) is set by distance-based criteria, while here it is the instantaneous set of Voronoi neighbors of particle $i$ at time $t$ \cite{ginelli2010relevance,sussman2021non}.
We work with the fixed value $\alpha=1/6$ -- in a periodic domain a generic tessellation will have six neighbors per particle -- and note that this fact gives us a properly extensive Hamiltonian. 
Finally, $\eta$ is the strength of the Gaussian white noise $\zeta_i$. Unless specified otherwise, we simulate $N=1.28\times10^6$ particles with $v_0=2.0$, and conduct simulations at a time-step size of $dt = 0.005$. We note that the precise value of the phase boundary is sensitive to the time-step size, which will be of note when we compare with $dt=1$ simulations in Sec. \ref{sec:discreteTime}.
We further note that some works on topological Vicsek-like models adopt a convention~\cite{martin2024fluctuation} of fixing $v_0$ and varying $\rho$ -- since there is no natural length-scale set by the particle interaction range these conventions are equivalent, but since we are adopting a time-continuous model we find it more convenient to fix the density.
In order to explore these system sizes in a computationally accessible manner, we adapt GPU-accelerated code from the \emph{cellGPU} package \cite{sussman2017cellgpu} to the model studied here.

To probe the nature of the phase transition in the above model, we focus much of our attention on the Binder cumulant~\cite{binder1981finite,binder1984finite,binder1987theory,binder1992monte}.
For an arbitrary random variable $f$ this is defined by
\begin{equation}
\label{eq::binderCumulant}
G_{f} = 1 - \frac{\langle{f^{4}}\rangle}{3\langle{f^2}\rangle^2}\,,
\end{equation}
where $\langle{f^2}\rangle$ and $\langle{f^4}\rangle$ denote the second and fourth moments of $f$ respectively.
Typically, the Binder cumulant of the global polar order parameter,
\begin{equation}
\phi(t)=\left|\frac{1}{N}\sum_{i=1}^N\mathbf{v}_{i}(t)\right|,
\end{equation}
is studied, and the averages in Eq.~\ref{eq::binderCumulant} are done over time \cite{chate2008collective,ginelli2010relevance}.
For a continuous transition, $G_{\phi}$ monotonically increases below the critical point as the mean of $\phi$ increases, while for discontinuous transitions $G_{\phi}$ will be non-monotonic in the vicinity of the phase transition (indicating a bimodal distribution of $\phi$).
In the context of these flocking models, the statistical expectation is that sampling $\phi(\mathbf{r},t)$ in disordered states is equivalent to sampling the norm of a collection of random unit vectors in 2D, which consequently yields $G_{\phi}=1/3$. In ordered states, the mean of $\phi(\mathbf{r},t)$ is non-zero mean and one finds that $G_{\phi}=2/3$ \cite{landau2021guide}.

In our work, we use large-enough systems that instead of looking at the statistics of $\phi(t)$ we can directly coarse-grain the velocity field of our system,
\begin{equation}
\mathbf{v}(\mathbf{r},t) = \frac{1}{\rho(\mathbf{r},t)}\sum_{i=1}^{N}\mathbf{v}_{i}(t)\delta\left[\mathbf{r}-\mathbf{r}_{i}(t)\right]\,,
\end{equation}
and the associated order parameter field
\begin{equation}
\phi(\mathbf{r},t) = \frac{\left|\sum_{i=1}^{N}\mathbf{v}_{i}(t)\delta\left[\mathbf{r}-\mathbf{r}_{i}(t)\right]\right|}{\sum_{i=1}^{N}\delta\left[\mathbf{r}-\mathbf{r}_{i}(t)\right]}\,.
\end{equation}
Definitionally, $\phi(t)$ and $\phi(\mathbf{r},t)$ are related by $\phi(t) = \int d^{2}\mathbf{r}\,\phi(\mathbf{r},t)\,$.
By sampling within sub-volumes of our system containing a sufficiently large number of particles, the statistics of $G_{\phi}$ will be the same as those generated by $\phi(t)$.
We note, though, that especially close to a critical point (when spatial correlation lengths may be large) analyzing the coarse-grained fields requires a sometimes delicate choice of the scale over which spatial averaging is done \cite{golden2023physically,gurevich2021learning}.
Figure \ref{fig:CoarseGrainedFieldSnapshots} shows representative coarse-graining scales applied to the same configuration of a $\sim$two-million particle configuration close to the order-disorder transition.
Readily apparent is that large averaging windows are needed to see some characteristic features. 
This fact will be further reflected in our analysis of the actual statistics and correlations of the resulting fields.

We also consider the statistics of the coarse-grained density field
\begin{equation}
\rho(\mathbf{r},t) = \sum_{i=1}^{N}\delta\left[\mathbf{r}-\mathbf{r}_{i}(t)\right],
\end{equation}
as was done in Ref.\cite{binder2021phase}.
In this case the expectation is $G_{\rho}=2/3$ in both the ordered and disordered phase, since density fluctuations about a homogeneous steady-state will be normally distributed about $\rho_{0}$.
If an inhomogeneous state exists, which here we expect to involve the presence of propagating flocks in high-density bands, then both $\rho(\mathbf{r},t)$ and $\phi(\mathbf{r},t)$ should be bimodal.
Consequently, $G_{\rho}$ and $G_{\phi}$ should exhibit minima in the vicinity of the transition \cite{chate2008collective} if the transition is discontinuous.

We will find it convenient to work in coordinates that reflect the direction of global polar order. When the polar order parameter is non-zero, we define $r_{||}$ and $r_{\perp}$ to be the direction parallel and transverse to the mean flocking direction, respectively.
With that convention, we follow Ref.~\cite{chate2008collective} and quantify the presence of transverse propagating bands by considering the longitudinal profile of a field $f$ averaged along $r_{\perp}$,
\begin{equation}
\label{eq::BandingProfile}
P_{f}(r_{||},t) \equiv \langle{f}(\mathbf{r},t)\rangle_{r_{\perp}}\,.
\end{equation}
From here, a ``banding order parameter'' is defined as
\begin{equation}
\label{eq::BandingOrderParameter}
B_{f}(t) \equiv \left\langle{P}_{f}(r_{||},t)^{2}-\left\langle{P}_{f}(r_{||},t)\right\rangle_{r_{||}}^{2}\right\rangle_{r_{||}}\,.
\end{equation}
This order parameter vanishes in both the disordered phase (for which we take $r_{||}$ and $r_{\perp}$ to be any orthogonal frame) and in the homogeneous polar flocking state.

\section{Coarse-grained field statistics}
\label{sec::CoarseGrainedFieldStatistics}

\begin{figure}[h]
	\centering
	\includegraphics[width=0.48\columnwidth]{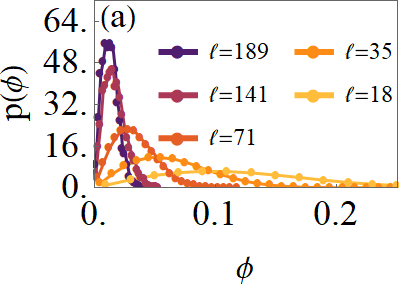}
	\includegraphics[width=0.50\columnwidth] {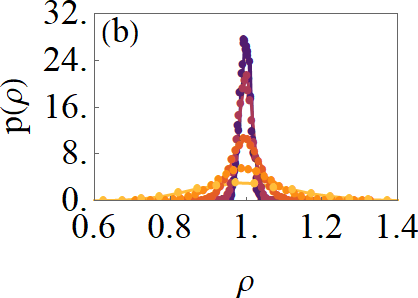}\vspace{4mm}
	\includegraphics[width=0.48\columnwidth]{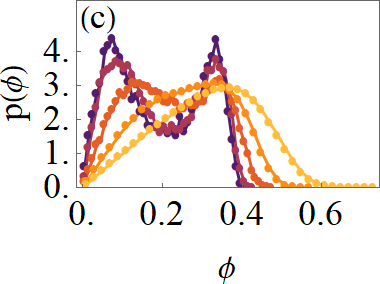}
	\includegraphics[width=0.50\columnwidth]{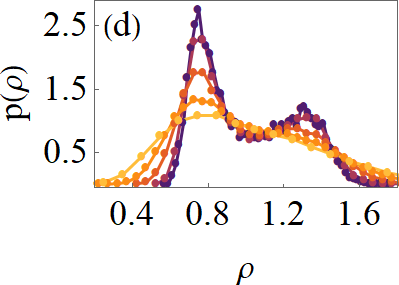}\vspace{4mm}
	\includegraphics[width=0.48\columnwidth]{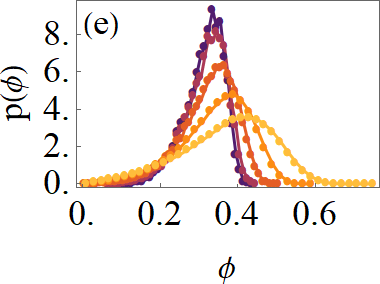}
	\includegraphics[width=0.50\columnwidth]{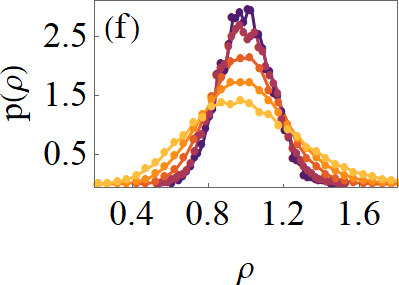}\vspace{4mm}
	\includegraphics[width=0.47\columnwidth]{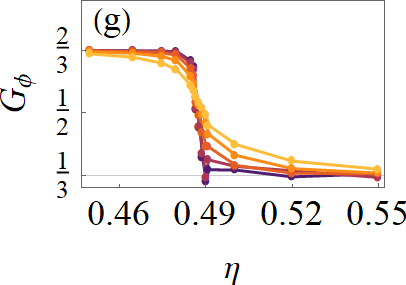}
	\includegraphics[width=0.51\columnwidth]{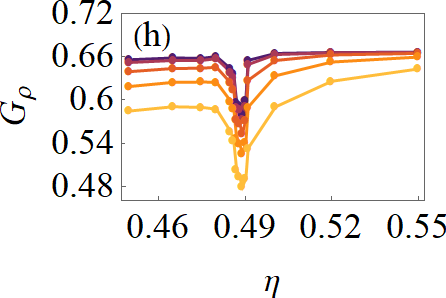}
	\caption{\textbf{Coarse-grained field statistics at different state points}.
        Figures (a-f) show the PDFs of local values of the order parameter and density field measured at different noise strengths and with coarse-graining length scales $\ell$ which increase from light to dark, as shown in the inset of (a).
        Figures (a,b) show data in the disordered phase ($\eta=0.5500$), figures (c,d) in the banding phase ($\eta=0.4890$) and figures (e,f) in the polar flocking phase ($\eta=0.4850$).
		The bottom row plots the Binder cumulant of the order parameter (g) and density field (h) measured in coarse-graining boxes of size $\ell$ as a function of noise strength.
	}
	\label{fig:CoarseGrainedFieldStatistics_Continuous}
\end{figure}

We first discuss the statistics of our model (Eqs.~\ref{eq::microscopic_pos_update}-\ref{eq::microscopic_Hamiltonian}) in the time continuous limit.
We varied the time-step size in our simulations between $dt =1.0$ and $dt=10^{-4}$, and established that our choice of $dt = 5\times 10^{-3}$ leads to convergence of all quantities of interest.
In Fig.~\ref{fig:CoarseGrainedFieldStatistics_Continuous} the probability density functions (PDFs) obtained from measurements of the distribution of local values of $\phi(\mathbf{r},t)$ and $\rho(\mathbf{r},t)$ are shown. 
Starting in the disordered phase (Figs.~\ref{fig:CoarseGrainedFieldStatistics_Continuous}a,b), the density field is symmetrically distributed about $\rho_{0}$, while the order parameter field distribution has a mean that vanishes as the coarse-graining length scale is increased.
Decreasing the noise strength so that the system is in the vicinity of the phase boundary (Figs.~\ref{fig:CoarseGrainedFieldStatistics_Continuous}c,d), we find that the distributions of density and order parameter values become non-Gaussian: they are skewed at small coarse-graining length scales and clearly bimodal at large length scales.
This is a key signature of phase separation in flocking models where low-density regions of space remain disordered, while high-density regions become ordered. 
Further decreasing the noise strength, Gaussian statistics of $\phi(\mathbf{r},t)$ and $\rho(\mathbf{r},t)$ are restored as the system enters a homogeneous flocking state (Figs.~\ref{fig:CoarseGrainedFieldStatistics_Continuous}e,f).

The deviations from Gaussian statistics in the vicinity of the phase boundary are quantified by the Binder cumulants $G_{\phi}$ and $G_{\rho}$ (defined by Eq.~\ref{eq::binderCumulant}) in Figs.~\ref{fig:CoarseGrainedFieldStatistics_Continuous}g,h.
In Fig.~\ref{fig:CoarseGrainedFieldStatistics_Continuous}g $G_{\phi}$ varies from $1/3$ (in the disordered phase) to $2/3$ (in the ordered phase) -- as found in earlier work on the Voronoi-Vicsek model \cite{chate2008collective,ginelli2010relevance} -- but does exhibit a very small dip below $1/3$ near the phase boundary, characteristic of a discontinuous transition.
This signal is significantly stronger in $G_{\rho}$, which exhibits a clear range of noise strengths for which the density field statistics are non-Gaussian.
From these results we conclude that the order-disorder transition of the spatially and temporally continuous Voronoi flocking model is indeed discontinuous.

\begin{figure}[h]
	\centering
	\includegraphics[width=0.51\columnwidth]{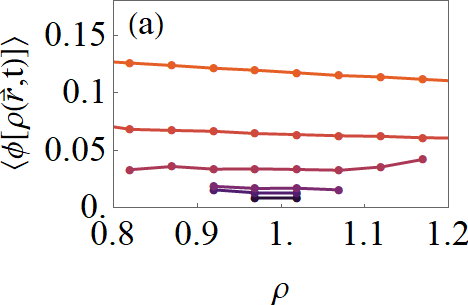}
	\includegraphics[width=0.47\columnwidth]{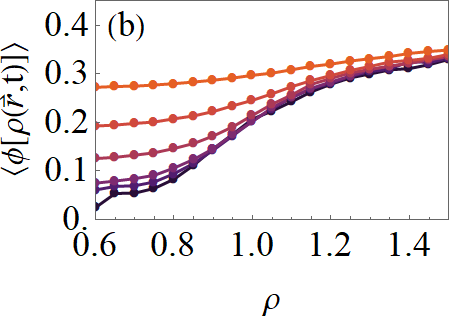}
	\caption{\textbf{Coupling between polarization and density fields}.
        The average local value of the order parameter field as a function of local density is shown in the disordered phase ($\eta=0.55$, left) and in the vicinity of the order-disorder transition ($\eta=0.4902$, right).
		Colors correspond to same coarse-graining bin sizes as in Fig.~\ref{fig:CoarseGrainedFieldStatistics_Continuous}.
	}
	\label{fig:PolarizationDensityCoupling_Continuous}
\end{figure}

Having found phase coexistence between the homogeneous disordered and ordered phases in the time-continuous limit of our topological model, we next investigate correlations between the order parameter and density fields.
It is this feature in the theoretically-predicted fluctuation-induced-first-order-transition scenario that renders homogeneous states near the phase boundary unstable.
In the disordered phase of our model, we find that the order parameter and density field are independent (Fig.~\ref{fig:PolarizationDensityCoupling_Continuous}a), as expected for a system of topologically interacting agents \cite{chou2012kinetic}.
In the vicinity of the phase boundary though, we measure a positive correlation between local order and local density, which \emph{increases} in strength with coarse-graining bin size (Fig.~\ref{fig:PolarizationDensityCoupling_Continuous}b) -- again emphasizing the need for the extremely large simulations employed in this work.
This observation confirms that the theoretically predicted hydrodynamic mechanism for producing banded states\cite{martin2021fluctuation} is indeed present in our simulations.

\section{Structure of the inhomogeneous phase}

\begin{figure}[h]
	\centering
	\includegraphics[width=0.49\columnwidth]{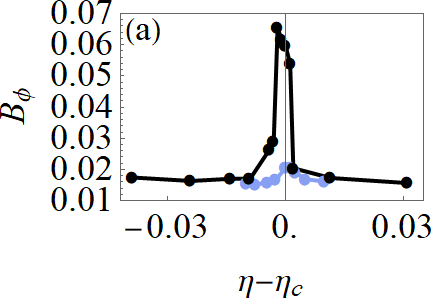}
	\includegraphics[width=0.49\columnwidth]{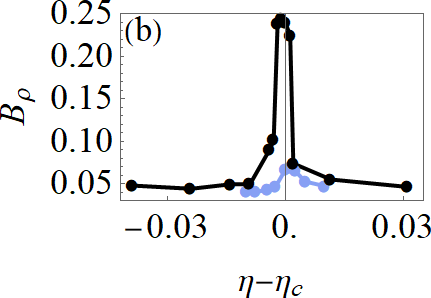}
	\includegraphics[width=0.49\columnwidth]{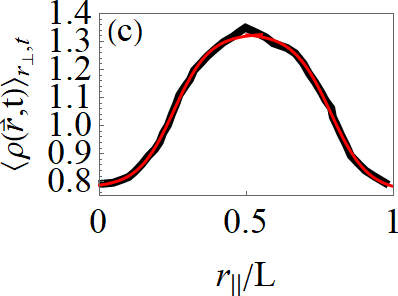}
	\includegraphics[width=0.49\columnwidth]{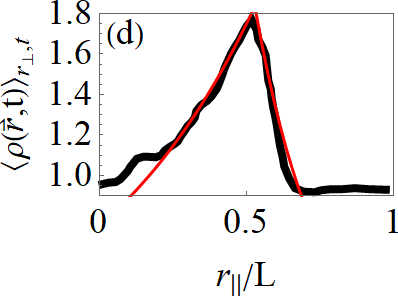}
	\caption{\textbf{Traveling band profiles in the phase coexistence regime}.
		The time-averaged value of the banding order parameter (Eq.~\ref{eq::BandingOrderParameter}) for the order parameter (a) and density fields (b) are shown for $dt = 0.005$ (black) and $dt=1.000$ (light blue).
		The time-average shape of the density field's profile in the vicinity of the phase boundary is shown for $dt=0.005$ and $v=2.0$ in (c) and $dt=0.010$ and $v=0.5$ in (d).
	}
	\label{fig:BandingProfiles}
\end{figure}

We next characterize the spatial structure of the observed states in the phase coexistence regime.
In Fig.~\ref{fig:BandingProfiles}a,b we show that highly dense and ordered, transversely extended bands exist in the vicinity of the phase boundary for the continuous time version of our model, whereas no coherent structures seem to form in the discrete time limit.
The precise structure of these bands is found by averaging data across many snapshots, translating the peak of the local transversely averaged density field to the center of the box.
As shown in Fig.~\ref{fig:BandingProfiles}c, these bands indeed have a phase separated profile predicted by field theory \cite{solon2015pattern}, with approximately half the system having $\rho(\mathbf{r},t)<\rho_{0}$ and the other half having $\rho(\mathbf{r},t)>\rho_{0}$.
In Fig.~\ref{fig:BandingProfiles}d we show that in different regions of phase space phase coexistence can take the form of asymmetric traveling bands, as is also permitted by the field theory \cite{solon2015pattern} and observed in metric flocks \cite{chate2008collective}.
We note that in all of our simulations we have only ever observed a single stable band, rather than the multiple bands that extremely large metric Vicsek systems can support.
We speculate that the intrinsic lack of a spatial length scale of interactions may result in only the longest-wavelength bands being stable.

\begin{figure}[h]
	\centering
	\includegraphics[width=0.70\columnwidth]{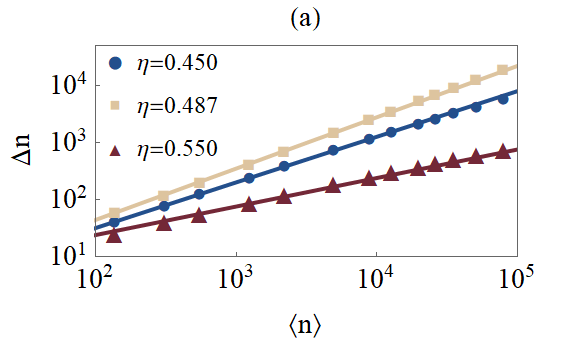}
	\includegraphics[width=0.44\columnwidth]{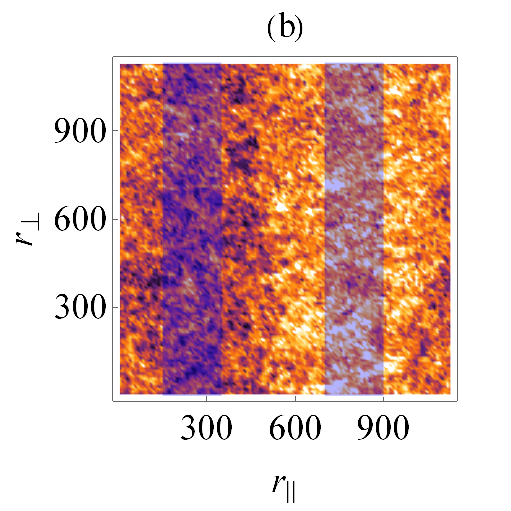}
	\includegraphics[width=0.44\columnwidth]{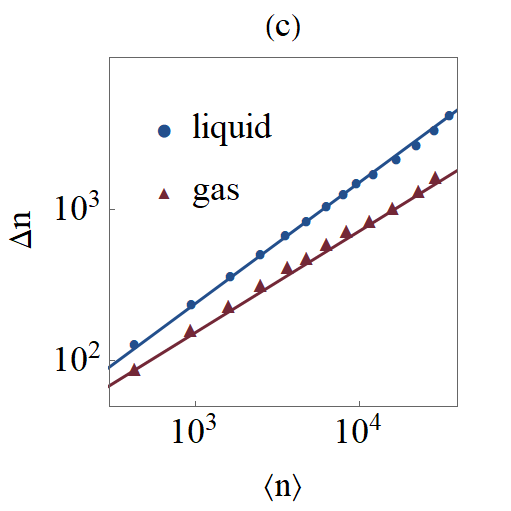}
	\caption{\textbf{Giant number fluctuations}.
    (a) Number fluctuations for a system in the disordered phase ($\eta=0.550$), the banded phase ($\eta=0.487$), and the ordered phase ($\eta=0.450$).
    Solid lines denote power law fits $\Delta{n}\propto\langle{n}\rangle^{\alpha}$ where $\alpha=0.5,0.9,0.8$ respectively.
    (b) A snapshot of a system in the banded phase, with superimposed blue rectangles denoting sub-volume regions of space in the disordered (gaseous) phase and the ordered (liquid phase).
    (c) Scaling of the number fluctuations with the sub-volume regions indicated in (b).
    Solid lines denoting power law fits with exponents $\alpha=0.80$ in the liquid region and $\alpha=0.67$ in the gaseous region.}
	\label{fig:giantNumberFluctuations}
\end{figure}

As another demonstration that the phase coexistence we see in our simulations is consistent with the known ordered and disordered states, we study the statistics of the density field within sub-volumes inside and outside of the bands.
Theoretical work predicts that the standard deviation of the number of particles within a region of space should scale as $\Delta{n}\propto\langle{n}\rangle^{\alpha}$, where $\langle{n}\rangle$ is the expected number of particles from the mean density of the system. 
The exponent $\alpha$ has the value $0.5$ in the disordered (gaseous) phase and $0.8$ in the ordered (liquid) phase where ``giant number fluctuations'' are present \cite{toner2005hydrodynamics} -- this is one of the key signatures of the polar flocking phase.
In Fig.~\ref{fig:giantNumberFluctuations}a, we show that our simulations reproduce these statistics, while also exhibiting an apparent third scaling regime in the inhomogeneous banded phase with an exponent close to $\alpha=0.9$.
As depicted in Fig.~\ref{fig:giantNumberFluctuations}b though, this additional regime goes away when carefully considering number fluctuations within and outside of the band.
When measured only in these spatial regions, we instead find the banded phase has a scaling exponent of $0.8$ inside the band and an exponent of $\sim0.67$ in the disordered ``gas'' outside the band (Fig.~\ref{fig:giantNumberFluctuations}c).
The former is consistent with the value measured in the homogeneous ordered phase, while the latter is identical to the value we measure for systems still in the disordered phase but very close to the phase transtion.

\section{Discrete time limit}
\label{sec:discreteTime}

\begin{figure}[b]
	\centering
	\includegraphics[width=0.470\columnwidth]{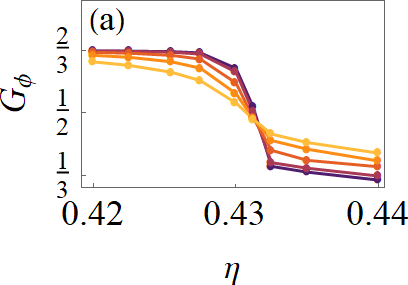}
	\includegraphics[width=0.510\columnwidth]{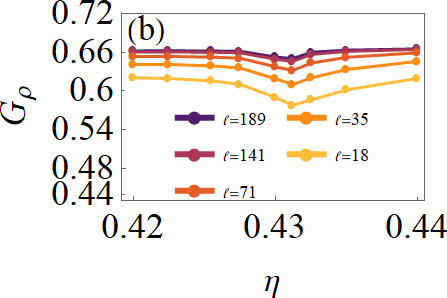}
	\caption{\textbf{Binder cumulant analysis in the discrete time limit}.
		The Binder cumulant of the order parameter (a) and density field (b) measured in coarse-graining boxes of size $\ell$ is shown for simulations with $dt=1.000$.
		The y-axis range in (b) has been made the same as in  Fig.~\ref{fig:CoarseGrainedFieldStatistics_Continuous}h for easier comparison}.
	\label{fig:CoarseGrainedFieldStatistics_Discrete}
\end{figure}

As previously mentioned, and shown in Fig.~\ref{fig:BandingProfiles}a,b, we are not able to observe propagating high-density bands in the discrete time limit ($dt = 1$) in which Vicsek models are traditionally simulated.
We further probe the apparent discrepancy between the finite-size scaling analysis from Ref.\cite{ginelli2010relevance} (which found no signature of a first-order transition), and our results in Section \ref{sec::CoarseGrainedFieldStatistics}.
One difference noted above is our use of a model with time-continuous dynamics (Eqs.~\ref{eq::microscopic_pos_update}-\ref{eq::microscopic_Hamiltonian}) as opposed to the discrete-time orientational dynamics of the Voronoi-Vicsek model studied in Ref.\cite{ginelli2010relevance}.
In Fig.~\ref{fig:CoarseGrainedFieldStatistics_Discrete} then, we compute the Binder cumulant of our model in the limit $dt=1.0$ just as was done in Fig.~\ref{fig:CoarseGrainedFieldStatistics_Continuous}g,h.
In this case we obtain the \emph{same} result as in Ref.\cite{ginelli2010relevance}, with $G_{\phi}$ varying smoothly and monotonically from $1/3$ to $2/3$.
For $G_{\rho}$, the dip previously observed in Fig.~\ref{fig:CoarseGrainedFieldStatistics_Continuous}h is significantly diminished, becoming almost indistinguishable for large coarse-graining sizes.
This indicates that in this parameter regime the large-scale density fluctuations in the system remain Gaussian even near the phase boundary.

\begin{figure}[h]
	\centering
	\includegraphics[width=0.50\columnwidth]{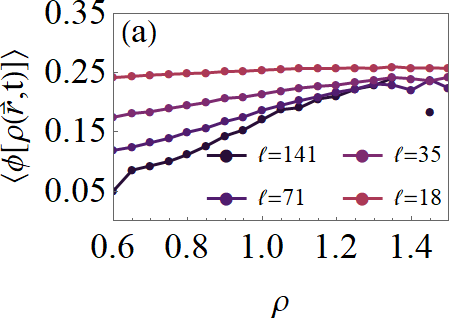}
	\includegraphics[width=0.48\columnwidth]{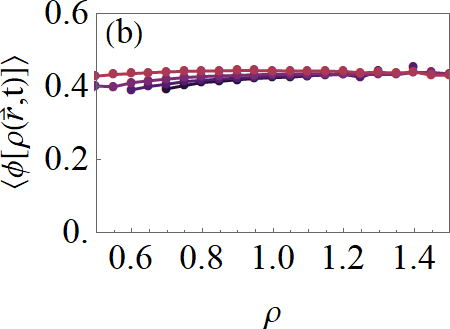}
    \caption{\textbf{Polarization-density coupling in the discrete-time limit}.
		The average local value of the coarse-grained order parameter field as a function of local density is shown for $dt=1.000$ in the vicinity of the phase boundary with $\eta=0.4312$ (left) and firmly in the homogeneous ordered phase with $\eta=0.4200$ (right).
	}
	\label{fig:PolarizationDensityCoupling}
\end{figure}

We also investigate how the discrete-time limit affects our observation of the fluctuation-induced coupling between the order parameter and density fields predicted by Ref.~\cite{martin2021fluctuation}.
As shown by the lightest curves in Fig.~\ref{fig:PolarizationDensityCoupling}, when measuring the local polarization as a function of local density with approximately the same $(\ell=16)$ coarse-graining bin size used in Ref.\cite{ginelli2010relevance}, we find no correlations between the two fields in the vicinity of the phase boundary or in the homogeneous ordered phase. 
As the coarse-graining bin size is increased though, we measure a strong and positive correlation near the phase boundary (while the fields remain largely independent in the homogeneous ordered phase).
We conclude that the fluctuation-induced coupling is sensitive to both finite-size effects and the noise strength in the system. 
The fact that a coupling of the two fields is indeed measured near the phase boundary (Fig.~\ref{fig:PolarizationDensityCoupling}a) suggests the hydrodynamic mechanism required to produce a linear instability and propagating bands are satisfied even in the discrete time limit. 
Although no discontinuity in the phase transition is observed in Fig.~\ref{fig:CoarseGrainedFieldStatistics_Discrete} at the system size we employ here, in the standard metric Vicsek model decreasing $dt$ is understood to greatly increase the system-size scale needed to observe the true discontinuous character of the transition \cite{chate2008collective}.
We speculate, then, that it is likely that the phase transition is discontinuous in the thermodynamic limit even in the discrete-time models of polar flocking.

\section{Discussion and outlook}

We have shown that the phase transition to collective motion in a Voronoi-Vicsek model with time-continuous orientational dynamics is discontinuous.
It displays clear phase coexistence between disordered and polar flocking states in a narrow but well-defined regime of parameter space, and further shows the coupling between density and order-parameter fields which is predicted to be a natural mechanism that leads to a discontinuous transition \cite{martin2021fluctuation}. 
We have further shown that the structure of the inhomogeneous banded states we observe is consistent with other field-theoretic predictions \cite{solon2015pattern}.

By investigating the stability of the banded states as a function of the discretization of time we provide a potential explanation for why previous finite-size analyses found only continuous transitions in Voronoi-Vicsek models \cite{ginelli2010relevance}.
Our work also suggests that anomalous results obtained in other numerical studies of flocking models -- such as metric variants of the model considered here having a continuous transition \cite{chepizhko2021revisiting} -- may also be due to a coarse discretization of time.
This warrants further work, as to the best of our knowledge there has not yet been a systematic study of how phase boundaries and the formation of mesoscopic structures in flocking systems depend on the size of the microscopic update time-step.

An open question in these topological models concerns the allowed spatial structure of the inhomogeneous phase.
As mentioned above, unlike in metric models we have only ever observed a single stable traveling band.
Is it possible for these topological models to support multiple bands traveling in the same direction? What about the stability of even more exotic phases, such as the ``cross-sea phase'' observed in metric flocking models, which have a pattern characterized by two non-parallel wavevectors \cite{kursten2020dry}?
At present there does not exist a hydrodynamic theory for understanding the instability that leads to the cross-sea phase, and further numerical work is required to probe the full extent of pattern formation in topological flocking models.



\section*{Conflicts of interest}
There are no conflicts to declare.

\section*{Data availability}
Data for this article, including representative simulation configurations and hydrodynamic fields used to generate the figures, are available at \href{zenodo.org}{[zenodo reference to be inserted]} The code used to run GPU-accelerated Voronoi-Vicsek simulations can be found at \href{https://github.com/sussmanLab/topologicalFlocking}{the sussmanLab github repository}\cite{sussman_2024_13646481}.

\section*{Acknowledgements}
We thank Julien Tailleur and Helen Ansell for helpful conversations. 
This material is based upon work supported by the National Science Foundation under Grant No. DMR-2143815.


\bibliography{topologicalFlockingBandedPhase} 
\bibliographystyle{rsc} 

\end{document}